\documentstyle[preprint,aps]{revtex}
\begin{document}
\draft
\title{CHAOS AND ENERGY REDISTRIBUTION IN THE 
NONLINEAR INTERACTION OF TWO SPATIO-TEMPORAL WAVE 
TRIPLETS\footnote{\scriptsize{PACS: 05.45.+b; Keywords: Nonlinear 
Wave Interaction, Hamiltonian Chaos, Spatio-Temporal Chaos, Nonlinear 
Dynamics}}}
\author{S.R. Lopes$^a$ and F.B. Rizzato$^b$\footnote{\scriptsize{Telephone: 
++ 55-51-3166470; Fax: ++ 55-51-3191762; E-mail: rizzato@if.ufrgs.br}}}
\address{$^a$Departamento de F\'{\i}sica - Universidade Federal do Paran\'a, 
P.O. Box 19081, 81531-990 Curitiba, Paran\'a, Brazil}
\address{$^b$Instituto de F\'{\i}sica - Universidade Federal do Rio Grande 
do Sul, P.O. Box 15051, 91501-970 Porto Alegre, Rio Grande do Sul, Brazil}
\maketitle
\date{\today}
\begin{abstract}
\noindent In this paper we examine the spatio-temporal dynamics of two 
nonlinearly coupled wave triplets sharing two common modes. Our 
basic findings are the following. When spatial dependence is absent, the 
homogeneous manifold so obtained can be chaotic or regular. 
If chaotic, it 
drives energy diffusion from long to small wavelengths as soon as 
inhomogeneous perturbations are added to the system. 
If regular, one may yet have two distinct cases: (i) energy diffusion 
is again present if the inhomogeneous modes are linearly unstable and 
triplets are effectively coupled; (ii) energy diffusion is absent if the 
inhomogeneous modes are linearly stable or the triplets are uncoupled. 
\end{abstract}


 
\newpage
\section{Introduction}

Wave interaction in continuous media is frequently described in terms of 
amplitude equations. The procedure used to obtain these amplitude equations is 
standard. Once one identifies a set of high-frequency modes interacting in a 
resonant fashion, one applies multiple scale formalisms to derive 
equations governing the slow space-time evolution of these modes; the 
amplitudes equations.

One of the most investigated types of resonant interaction that can 
be effectively described by amplitude equations corresponds 
to the triplet interaction. In that case, three high-frequency 
modes undergo a resonant interaction which, in the absence of 
external drives and dissipation, can be shown to produce integrable 
space-time dependent amplitude equations \cite{kau79,wil77,chow95}. 

Another case of resonant interaction that has been attracting some 
renewed attention is the one involving two coupled triplets sharing 
two common modes \cite{sug68,kar73,wal77,chi96,pak97}. This type of 
interaction has been shown to produce amplitude equations allowing for 
temporal chaos \cite{chi96,pak97}. 
To be more specific, if one removes all spatial dependence from the governing 
equations, then the solutions developing on this {\it homogeneous manifold} 
can be chaotic if a number of conditions are fulfilled. Not much, 
however, has been said about solutions departing from the homogeneous 
manifold. One relevant question here would be on the stability of 
the homogeneous manifold against inhomogeneous perturbations. If, 
for instance, a certain homogeneous solution - not necessarily a static 
equilibrium - is perturbed with some small inhomogeneity, would this 
inhomogeneity grow in time? If the answer is positive we would have an 
{\it unstable homogeneous manifold}. Otherwise the manifold would be 
{\it stable} or {\it marginally stable}. One could actually split the 
stability analysis into the analysis of homogeneous manifolds supporting 
regular dynamics and homogeneous manifolds supporting chaotic dynamics, and 
the respective results could be significantly different. In fact, as 
we have observed and shall discuss in detail later, chaotic manifolds are 
intrinsically unstable. In other words, while regular manifolds may be unstable 
against some types of perturbation and stable against others, chaotic manifolds 
always tend to be unstable no matter the type or strength of the perturbation.
What happens in the latter case is that inhomogeneities always tend to grow as a 
result of the stochastic drive provided by the chaotic orbits evolving on the 
homogeneous manifold. Here one cannot use linear techniques to investigate 
stability since the problem is intrinsically nonlinear; one should make use of 
stochastic models \cite{lili83} instead. On the other hand, if homogeneous chaos 
is absent the stochastic drive is absent as well. Then stability can be estimated 
by linear analysis which in fact shows that only modes falling within the linear 
instability band are the ones to grow. 

Based on these comments one can already see that 
at least in the case of chaotic homogeneous manifolds, the instability is 
unlikely to be saturated by any nonlinear effect involving a 
small number of modes, like soliton formation for instance. As energy is 
continuously fed into the inhomogeneities, modes with smaller and smaller 
length scales are eventually excited. Interestingly, we shall also see that in 
the case of regular homogeneous manifolds coupled to linearly unstable modes, 
the behavior is similar; the initial instability is not saturated by nonlinear 
effects involving a finite number of modes and once more energy spreads toward 
smaller and smaller length scales. Energy flow is arrested only when the 
triplets are uncoupled, or else when the homogeneous manifold is regular and no 
inhomogeneous mode falls within the linear instability band.   

In view of the previous facts, the issues to be discussed here are 
related to topics on turbulent motion or on the Fermi-Pasta-Ulam problem 
\cite{pet90,goe92,tsa96}. Indeed, our homogeneous manifold could be seen as 
representing fluctuations with very long wavelengths. Therefore the stability 
problem that we wish to pose here could be more precisely stated as 
follows: when is our system unstable and under which conditions can energy 
flow from long to short wavelengths?  Contrarily to integrable cases where no 
flow is observed, we shall see here that {\it energy transfer} or {\it 
energy redistribution} does occur under the conditions mentioned above: 
(i) when the homogeneous manifold is chaotic; or 
else (ii) even with a regular homogeneous manifold, when the two triplets are 
effectively coupled and the inhomogeneous perturbations fall within the 
instability band.

The paper is organized as follows: in \S II we introduce the basic 
formalism, governing equations, and numerical methodology; in \S III 
we analyze the dynamics on the homogeneous manifold with help of Poincar\'e 
maps, in \S IV we perform the relevant analyses and simulations for the full 
spatio-temporal problem, and in \S V we conclude the work. 

\section{The Model}

We are interested in describing the mutual interaction of two wave 
triplets sharing two common modes and constituted, therefore, by four wave 
modes in all. We assume that the following resonant conditions are 
fulfilled:
$$
\omega_{\kappa_1} \approx \omega_{\kappa_3}+\omega_{\kappa_2} ,$$
$$
\kappa_1=\kappa_2+\kappa_3,$$
$$
\omega_{\kappa_1} \approx \omega_{\kappa_4}-\omega_{\kappa_2},$$
\begin{equation}
\kappa_1=\kappa_4-\kappa_2,
\label{quadrupleto}
\end{equation}
where $\omega_{\kappa_j}$ and $\kappa_j$, $\{j=1,2,3,4\}$, are respectively 
the fast frequencies and fast wavevectors of the interacting waves. 

Note that we allow for small frequency mismatches, an usual effect in wave-wave 
interaction. Frequency mismatch occurs because even for perfectly matched 
wavevectors, the respective frequencies obtained from the linear dispersion 
relations may not be quite likewise matched. In laser accelerators and 
in general laser-plasma interactions, frequency mismatch can even 
enhance the linear growth rate in several situations \cite{chi96,riz92}. 
Therefore, mismatched modes can be stronger and of greater importance in the 
dynamics. Moreover, as we shall see here, the integrability properties of the 
nonlinear regime of the interaction depend critically on the existence of mismatch. 
Wavevector mismatches could be also incorporated into the theory. However, as 
frequency and wavevector mismatches are formally equivalent we focus attention on 
the former. 

This type of kinematics where the interacting triplets share 
two common modes, $1$ and $2$, has been shown to be of relevance in nonlinear 
interactions involving electromagnetic, Langmuir and ion-acoustic waves propagating 
in unmagnetized plasmas \cite{sug68,chi96}, in magnetohydrodynamics interactions 
involving Alfv\'en and ion-acoustic waves \cite{kar73}, and in plasma-beam 
interactions in the presence of negative energy waves \cite{wal77}. 
Following the model revisited in a series of recent 
papers \cite{chi96,pak97}, the dimensionless amplitude equations governing one 
dimensional, spatio-temporal, slow modulational dynamics can be cast into 
the form: 
\begin{equation}
{\partial A_1(x,t) \over \partial t} + 
v_{g_1} \, {\partial A_1(x,t) \over \partial x} =
A_2(x,t) A_3(x,t) - r \> A_2^\ast (x,t) A_4(x,t),
\label{a1c}
\end{equation}
\begin{equation}
{\partial A_2(x,t) \over \partial t} + 
v_{g_2} \, {\partial A_2(x,t) \over \partial x} =
-A_1(x,t) A_3^\ast(x,t) - r \> A_1^\ast(x,t) A_4(x,t),
\label{a2c}
\end{equation}
\begin{equation}
{\partial A_3(x,t) \over \partial t} + 
v_{g_3} \, {\partial A_3(x,t) \over \partial x} =
i\delta_3 A_3(x,t) - A_1(x,t) A_2^\ast(x,t),
\label{a3c}
\end{equation}
\begin{equation}
{\partial A_4(x,t) \over \partial t} + 
v_{g_4} \, {\partial A_4(x,t) \over \partial x} =
i\delta_4 A_4(x,t) + r \> A_1(x,t) A_2(x,t). 
\label{a4c}
\end{equation}
$A_j$, $\{j=1,2,3,4\}$, are the 
complex amplitudes of the four fields, 
$\delta_{3,4} = \omega_{1} \mp \omega_{2}-\omega_{3,4}$ are independent 
frequency mismatches corresponding to fields $A_3$ and $A_4$ (i.e., one 
can always take $\delta_1=\delta_2=0$, as we actually did), $r$ is a 
variable strength factor measuring the intensity of the triplet-triplet 
coupling \cite{chi96,pak97}, and $v_{g_j}$ are the respective group 
velocities along the spatial modulation that we take as one dimensional
(the $x$ axis) in this work. Time and space derivatives are first order as a 
result of our multiple time and space scales. 

The set of governing equations (\ref{a1c})-(\ref{a4c}) can be derived 
from a continuous Hamiltonian. Indeed, it is possible to see that 
the following relations hold
\begin{equation}
{\partial A_j(x,t) \over \partial t} = {\delta H \over \delta A_j^\ast},\>\>
{\partial A_j(x,t)^\ast \over \partial t} = - {\delta H \over \delta A_j},
\label{canoni}
\end{equation}
where we introduce the functional derivative 
\begin{equation}
{\delta \over \delta A_j} \equiv {\partial \over \partial A_j} - 
{\partial \over \partial x} \>\> {\partial \over \partial 
({\partial A_j \over \partial x})}
\label{funcdev}
\end{equation}
(the same definition holds if $A$ is replaced with $A^\ast$), and 
where the Hamiltonian is to be written in the form 
$$
H = \int dx [ - A_1 A_2^\ast A_3^\ast + A_1^\ast A_2 A_3 - 
    r (A_1^\ast A_2^\ast A_4 - A_1 A_2 A_4^\ast) +
$$
\begin{equation}
    i \delta_3 |A_3|^2 + i \delta_4 |A_4|^2 - 
    \sum_{j=1}^4 \, v_{g_j} \, A_j^\ast {\partial A_j \over \partial x}].
\label{hamiltoniano}
\end{equation}
The Hamiltonian does not depend explicitly on time; therefore it is a time 
conserved quantity. In addition to the Hamiltonian, the following 
quantities are also conserved:  
\begin{equation}
C_1 = \int dx \left[ |A_2|^2 - |A_3|^2 + |A_4|^2 \right],
\label{c1}
\end{equation}
\begin{equation}
C_2 = \int dx \left[ |A_1|^2 + |A_3|^2 + |A_4|^2 \right], 
\label{c2}
\end{equation}
and they suggest that we can look at the whole process as a decay 
interaction with $A_1$ as the decaying pump; from this perspective, 
$A_2$ is an idler wave whereas $A_3$ is a Stokes mode and $A_4$ an 
anti-Stokes mode. In the case of nonlinear interactions in unmagnetized 
plasmas for instance, wave $A_1$ is a transverse electromagnetic wave, 
wave $A_2$ is ion-acoustic, and waves $A_3$ and $A_4$ are Stokes and 
anti-Stokes Langmuir modes \cite{sug68,chi96}. 

All those quantities will be useful in checking out the accuracy of our 
integration methods which we outline now. The basic integration method 
is pseudo-spectral and the four fields $A_j\,,\{j=1,2,3,4\}$, are
Fourier analyzed according to 
\begin{equation}
A_j = \sum_{n=-N/2+1}^{N/2} a_{j_n} (t) e^{i n k x}.
\label{fourier}
\end{equation}
For further purposes note that while $j$ denotes the {\it wave type}, $n$ 
denotes the {\it mode number} or {\it harmonic number}. The harmonic or modal 
amplitudes $a_{j_n}$ are integrated in time with a predictor-corrector 
algorithm. We use $N=64,\,128,\,256$ modes removing half of them to cure 
aliasing. We denote the basic slow wavevector by $k$ and point out that due 
to the structure of the equations, variations in $k$ can be absorbed in 
variations of the group velocities or vice-versa.  Fluctuations of the 
conserved quantities, including energy, are not larger than one part in 
$10^7$, and variations of the tolerance factors of the integrator do not alter 
the final outcome of the runs.  

\section{Dynamics on the Homogeneous Manifold}

As a first step let us analyze the behavior of the dynamics on the 
homogeneous manifold. The motivation and relevance of a consistent 
analysis of the homogeneous dynamics has been commented in the 
Introduction and has roots in the stochastic pump model that we 
discuss next. According to the stochastic pump model \cite{lili83}, 
as soon as strongly chaotic degrees-of-freedom of a dynamical system 
are coupled to additional degrees-of-freedom, the 
chaotic subsystem acts like a stochastic pump 
delivering a net amount of energy to its environment. 
This net energy transfer is diffusive 
and takes place at all because of the very random nature of the coupling. 
In a certain sense, the dynamics is equivalent to the coupling of two 
thermodynamical systems, one cold and the other hot. As a result of the random 
coupling, a net energy transfer finally equalizes the temperatures of both 
systems. In our case, the hot system would be the chaotic homogeneous manifold 
and the cold one the subset of the inhomogeneous modes. Furthermore, and 
precisely due to diffusion, energy would be redistributed over the accessible 
phase-space of the inhomogeneous modes exciting progressively harmonics with 
finer and finer length scales. 

Homogeneous orbits have been recently investigated in a paper by 
Pakter, Lopes, \& Viana \cite{pak97}. Let us review the basic 
results and take the opportunity to introduce a convenient notation to be 
used here. To investigate the types of orbits on the homogeneous manifold, let 
us thus assume that we are working with fields of the form
\begin{equation}
A_j(x,t) \equiv a_{j_o}(t),
\label{ajo}
\end{equation}
which do not depend on the spatial coordinate.
Then, introducing real amplitudes and phases through 
\begin{equation}
a_{j_o} = \sqrt{\rho_j} e^{i \phi_j},\> \{j=1,2,3,4\},
\label{introduzro}
\end{equation}
one obtains canonical conjugate equations for these quantities, with a 
reduced governing Hamiltonian $h$ given in the form
\begin{equation}
h = 2 \sqrt{\rho_1 \rho_2 \rho_3} \> \sin (\phi_1-\phi_2-\phi_3) - 
    2 r \sqrt{\rho_1 \rho_2 \rho_4} \> \sin (\phi_1+\phi_2-\phi_4)
    -\delta_3 \rho_3 - \delta_4 \rho_4.
\label{hreduz}
\end{equation}
This type of Hamiltonian structure has been largely studied in the context 
of nonlinear waves \cite{nos97}. What we learn is that due to the particular phase combinations of the sine functions, a number of relatively simple canonical 
transformations can be applied to simplify the problem. The first 
transformation introduces a new phase angle $\phi_1'$ in the form 
$\phi_1' = \phi_1-\phi_2-\phi_3$ 
and preserves the other phases, $\phi_j'=\phi_j$, $\{2,3,4\}$. Rewriting 
conveniently the argument of the second sine function of expression 
(\ref{hreduz}), the form of the second canonical 
transformation becomes clear; it shall be constructed such as to introduce a 
new phase angle 
$\phi_4'' = \phi_4' - 2 \phi_2' - \phi_3'$, preserving all the others 
$\phi_j''=\phi_j'$, $\{j=1,2,3\}$. As a result of the pair of transformations 
described above, one finally arrives at  
\begin{equation}
\rho_2 = \rho_2'' - \rho_1'' - 2 \rho_4'',
\label{rho2}
\end{equation}
\begin{equation}
\rho_3 = \rho_3'' - \rho_1'' - \rho_4'',
\label{rho3} 
\end{equation}
and  
$$
h= 2 \sqrt{\rho_1 (\rho_2 -\rho_1 - 2 \rho_4) (\rho_3-\rho_1 - \rho_4)} 
   \> \sin(\phi_1) - $$
\begin{equation}
  2 r \sqrt{\rho_1 \rho_4 (\rho_2 - \rho_1 - 2 \rho_4)} \> 
   \sin(\phi_1 - \phi_4) - 
   \delta_3 (\rho_3 - \rho_1 - \rho_4) - 
   \delta_4 \rho_4,
\label{hreduzlinha}
\end{equation}
where the double primes have been dropped in the Hamiltonian 
(\ref{hreduzlinha}). Since the Hamiltonian no 
longer depends on coordinates $\phi_2$ and $\phi_3$, the associated momenta 
$\rho_2$ and $\rho_3$ are constants of motion with their relation to the original 
variables given by relations (\ref{rho2}) and (\ref{rho3}). Furthermore, 
the Hamiltonian $h$, Eq. (\ref{hreduzlinha}), is a conserved quantity itself 
since it does not depend explicitly on time. We now focus attention on 
conditions allowing for decay processes where $A_1$ is to be considered a pump 
wave. In the decay case one requires $A_1 \neq 0$ and $A_2=A_3=A_4=0$, which 
implies, in view of conditions (\ref{ajo})-(\ref{hreduzlinha}), that our 
energy surface of interest here is actually indexed by $h = 0$. 

The flow governed by $h$ is two-degrees-of-freedom and 
thus likely to be nonintegrable \cite{lili83}. The relevant governing 
equations obtained from Hamiltonian (\ref{hreduzlinha}) are:

\[
{d \rho_1 \over dt} =
-2\sqrt{\rho_1(\rho_2-\rho_1-2\rho_4)(\rho_3-\rho_1-\rho_4)}\cos\phi_1
\]
\begin{equation}
+2r\sqrt{\rho_1\rho_4(\rho_2-\rho_1-2\rho_4)}\cos(\phi_1-\phi_4), 
\label{hreduz1}
\end{equation}
\begin{equation} 
{d \rho_4 \over dt} = - 2r\sqrt{\rho_1\rho_4(\rho_2-\rho_1-2\rho_4)}\cos(\phi_1-\phi_4), 
\label{hreduz2}
\end{equation}  
\[ 
{d \phi_1 \over dt} =
{1\over\sqrt{\rho_1(\rho_2-\rho_1-2\rho_4)(\rho_3-\rho_1-\rho_4)}}\sin\phi_1
\times
\]
\[ 
\left
[ \rho_2\rho_3-2\rho_1\rho_2-\rho_2\rho_4-2\rho_1\rho_3+3\rho_1^2+6\rho_1\rho_4- 
2\rho_4\rho_3+2\rho_4^2\right ] -
\]
\begin{equation}
{r\sin(\phi_1-\phi_4)\over\sqrt{\rho_1\rho_4(\rho_2-\rho_1-2\rho_4)}} 
\left[\rho_4\rho_2-2\rho_1\rho_4-2\rho_4^2\right ] + \delta_3, 
\label{hreduz3} 
\end{equation}
and
\[ 
{d \phi_4 \over dt} =
{1\over\sqrt{\rho_1(\rho_2-\rho_1-2\rho_4)(\rho_3-\rho_1-\rho_4)}}\sin\phi_1
\times
\]
\[
\left[ -2\rho_1\rho_3+3\rho_1^2+4\rho_1\rho_4-\rho_1\rho_2\right]- 
\]
\begin{equation}
{r\sin(\phi_1-\phi_4)\over\sqrt{\rho_1\rho_4(\rho_2-\rho_1-2\rho_4)}} 
\left [ \rho_1\rho_2-\rho_1^2-4\rho_1\rho_4 \right ] +\delta_3 -\delta_4.
\label{hreduz4}
\end{equation}

However there is at 
least one pair of limiting cases where the dynamics becomes completely
integrable. One of them occurs when $\delta_3=\delta_4=0$. In this situation 
Romeiras \cite{rom83} has identified a new constant of motion that reduces the 
system to one-degree-of-freedom. The other integrable limit occurs when 
$r \rightarrow 0$; in this situation mode $A_4$ decouples from the 
multiplet and one recovers the pure three wave decay which, again, is 
one-degree-of-freedom and hence completely integrable. 

To understand and visualize the properties and characteristics of the purely 
homogeneous interaction, let us fix $r=1.0$ - a typical condition 
of interactions processing in unmagnetized plasmas or magnetohydrodynamics 
environments, for instance - and $\delta_4=0$. Let us then proceed to 
investigate the system behavior as $\delta_3$ is allowed to vary. We 
make use of Poincar\'e plots, where we record the values of 
$\rho_1$ and $\phi_1$, as determined by the governing reduced Hamiltonian 
(\ref{hreduzlinha}), each time $d\rho_4/dt = 0$ with 
$d^2 \rho_4 /dt^2 > 0$. We consider weak nonlinearities and launch 
a number of initial conditions, in fact $10$, under the constraint $h=0$. In 
all the conditions we take $\phi_1=\pi/2$, $\phi_4=0$; $\rho_1$ is uniformly 
distributed within a weakly nonlinear range $0<\rho_1<\rho_{max}<1$ and 
$\rho_4$ is calculated from $h=0$. Finally, to be consistent with typical 
decay conditions, we take $\rho_2 = \rho_3 = \rho_{max}$ such that 
whenever $\rho_1 = \rho_{max}$ and $\rho_4=0$, $A_2=A_3=0$. 

As mentioned, in this paper we investigate weakly nonlinear regimes; all 
the simulations are performed  with $\rho_{max}=0.1$. Now the first plot 
shown in Fig. (\ref{fig1.ps}a) is produced for parameters lying in an 
essentially regular regime where $r=1$, and $\delta_3=0.0001$. As predicted by 
Romeiras \cite{rom83}, for such a small $\delta_3$ the phase-space is 
essentially regular displaying a series of closed curves that in fact reveal 
the existence of numerous invariant KAM (Kolmogorov-Arnold-Moser) tori 
\cite{lili83}. Then, as we increase the value of the mismatch up to 
$\delta_3=0.1$, Fig. (\ref{fig1.ps}b) indicates that the KAM tori become 
substantially eroded; chaos is strong in this case. As we couple the 
homogeneous manifold to inhomogeneous perturbations, diffusion towards small 
wavelengths is expected to occur in the latter situation. 

We have therefore obtained some initial information on the behavior of the 
homogeneous manifold. But the real question yet to be answered is how that 
kind of behavior on this manifold affects the inhomogeneous dynamics. We 
shall address this issue next.

\section{Full Spatio-Temporal Dynamics}

The typical initial conditions to be used in this paper place the homogeneous 
orbits on the outermost curve of Fig. (\ref{fig1.ps}a), close to a separatrix orbit 
containing two unstable fixed points located at the maximum $\rho_1$ - in this case 
$\rho_1 = 0.1$ - and $\phi_1=0,\pi$. These fixed points represent the exact 
homogeneous equilibria of our system. In regular cases the orbit remains close 
to the separatrix and since the separatrix orbit spends an infinitely large amount 
of time close to the fixed points, we estimate linear stability simply by replacing 
the actual orbit with the fixed points. The estimates shall be shown to be accurate 
enough when compared with the simulations.  

Let us thus suppress for a little while the dynamics on the homogeneous manifold 
and assume that we have an equilibrium situation characterized by 
$A_1^{(0)} = a_{1_0}^{(0)} \equiv {\cal A} \neq 0$ 
(${\cal A}$ is a complex constant), and $a_{j_n}^{(0)}=0\,,\{j=2,3,4\}$, 
for any other $n$. As mentioned before, this is the convenient condition if 
one wishes to study the stability of a strong pump - mode $A_1$ - feeding 
vanishingly small modes - modes $A_j$, $\{j=2,3,4\}$. 
The stability of the pump can then be readily investigated if one adds 
small perturbations 
$A_j^{(1)} = a_{j_1}^{(1)} e^{i k x }, \,a_{j_1}^{(1)} \sim e^{-i \Omega t}$ 
to the system. Indeed, if 
$|a_{j_1}^{(1)}| \ll |A_1^{(0)}|$, a dispersion relation 
$\Omega = \Omega (k)$ can be obtained from
\begin{equation}
P_3 \, P_2-P_R=0,
\label{reldis}
\end{equation}
where
$$P_3=i (- \Omega + v_{g_3} k - \delta_3),$$ 
$$P_2=i (- \Omega + v_{g_2} k),$$ 
$$P_R=|{\cal A}|^2 (1-r^2 {P_3 \over P_4}),$$ 
$$P_4=i (- \Omega + v_{g_4} k - \delta_4).$$
The dispersion relation (\ref{reldis}) can be analyzed numerically but here we 
abbreviatedly discuss some of its relevant results. For a given 
${\cal A}$, there exists an instability (complex $\Omega$'s) band extending 
from $k=0$ up to a certain $k_{max}$. The larger the value of ${\cal A}$, 
the larger $k_{max}$, but otherwise the instability band is not strongly 
dependent on $\delta_3$, $\delta_4$, and $r$. The instability for $k = 0$, in 
particular, is the one giving rise to the homogeneous dynamics that was the 
subject of study in section II and in previous works \cite{chi96,pak97}. As a 
relevant information obtained there, we recall that the ensuing homogeneous 
dynamics can be chaotic or regular depending on the initial conditions and 
control parameters.

We can now collect the results obtained so far to assert that in general 
we have four global types of situations to be 
investigated. These four situations result from the four distinct associations 
involving a regular {\it or} chaotic homogeneous manifold, with 
inhomogeneous perturbations launched inside {\it or} outside the linear 
instability band. Our interest is to discover in which of the four 
situations can one detect energy diffusion from long to small wavelengths. 
Although all four cases are of interest, in this paper we shall focus on the two independent situations that provide a clear view on the individual 
role of the basic nonlinear effects driving diffusion: chaotic homogeneous manifolds associated with linearly stable inhomogeneous perturbations, and 
regular homogeneous manifolds associated with linearly unstable inhomogeneous 
perturbations. The behavior of the two remaining cases may be then 
obtained by an extension of the behavior of these basic ones. Our 
classification outlined above may sound artificial since the 
homogeneous and inhomogeneous dynamics are in fact coupled and 
mutually interactive; in other words they should not be considered 
apart from each other in principle. However, the classification adopted 
explains well enough what we are about to see: in the case of chaotic 
homogeneous manifolds coupled to stable modes, the diffusion is so slow that 
at any instant one can consider the chaotic dynamics as adiabatically 
delivering energy to the inhomogeneous modes; in the case of regular 
homogeneous manifolds coupled to unstable modes, similarly, a possible regular 
dynamics on the homogeneous manifold does not alter significantly the results 
of linear stability calculations.    

To monitor instabilities and energy transfer, we make use of a spectral average \cite{yue80,thy81} that enables to estimate the number of active modes in the 
system. We denote this quantity by $\sqrt{<N^2>}$ and define it as 
\begin{equation}
\sqrt{<N^2>} = \sqrt{{\sum_{n=-N/2+1}^{N/2} \sum_{j=1}^4 n^2 |a_{j_n}|^2 
\over \sum_{n=-N/2+1}^{N/2} \sum_{j=1}^4 |a_{j_n}|^2}}.
\label{speave}
\end{equation}
From its definition one sees that $\sqrt{<N^2>}$ is the square root of the averaged 
$n^2$. The average is taken over mode number $(n)$ and mode type $(j)$, and weighted 
by the square of the mode amplitude for a fixed initial condition. One could also average 
over several similar initial conditions but our results on energy transfer remain the same 
as long as the initial conditions are all simultaneously stable or all simultaneously 
unstable. $\sqrt{<N^2>}$ is expected to grow in time in diffusive cases where more and 
more modes become involved in the dynamics. In the absence of energy transfer, 
$\sqrt{<N^2>}$ remains limited by the number of linearly unstable modes; in 
the case of stable modes only, $\sqrt{<N^2>} \rightarrow 0$.

In our simulations we take $\delta_4=0$. Our results on energy transfer 
are qualitatively independent of a precise choice of group velocities. Therefore 
we consider $v_{g_j}=0,\,\{j=1,3,4\}$ for simplicity, since variations of 
$v_{g_2}$ alone provides an easy way to control the width of the band of 
linear instability. 
We point out that a precise choice of velocities would be essential either in 
the case of a pure triplet interaction, or else if one were performing a detailed 
study of how the dynamics of the four-wave system evolves in time during intermediary 
stages, as the asymptotic state is approached. Indeed, depending on the ordering of group 
velocities solitons can actually saturate the pure triplet interaction or perhaps can serve 
as metastable intermediary states existing for a finite amount of time when a fourth wave 
destroying integrability is added. However, once the four-wave system has a chaotic 
homogeneous manifold or becomes linearly unstable as discussed earlier, energy transfer 
progresses regardless the presence or absence of solitons in these intermediary states. 
Note that when the coupling parameter $r$ is sufficiently large, the dynamics of the 
four-wave system does not resemble that of the triplet interaction. This is why we adopt 
the simple ordering discussed above; other orderings not presented in the present work have 
been tested as well.  

In any case, such a choice is of physical significance for 
the interaction of electromagnetic, ion-acoustic and Langmuir waves in plasmas 
with small enough Debye length. In that case one shows that the only relevant 
group velocity is the one corresponding to the ion-acoustic mode, the $A_2$ 
wave \cite{chi96}. The choice of the numerical value for $v_{g_2}$ and $k$ 
is arbitrary to a large extent. This is due to the structure of the governing equations 
(\ref{a1c})-(\ref{a4c}), which are invariant under the rescalings 
\begin{equation}
v_{g_j} \rightarrow \alpha \, v_{g_j},\> x \rightarrow \alpha x 
\label{scale1}
\end{equation}
and
\begin{equation}
A_j \rightarrow \beta A_j,\> t \rightarrow {t \over \beta}, \> 
x \rightarrow {x \over \beta}, \> \delta_j \rightarrow \beta \delta_j, \> 
\{j=1,2,3,4\},
\label{scale2}
\end{equation}
$\alpha$ and $\beta$ being independent scale factors. This fact 
effectively creates some additional freedom in choosing the 
simulation parameters. Here we shall consider $k=1.0$ in all cases.
 
\subsection{The $\delta_3 \neq 0$ case.}
 
\begin{figure}
\caption{Poincar\'e plots on the homogeneous manifold for 
$r=1.0$: $\delta_3=0.0001$ in (a) and $\delta_3=0.1$ in (b)}
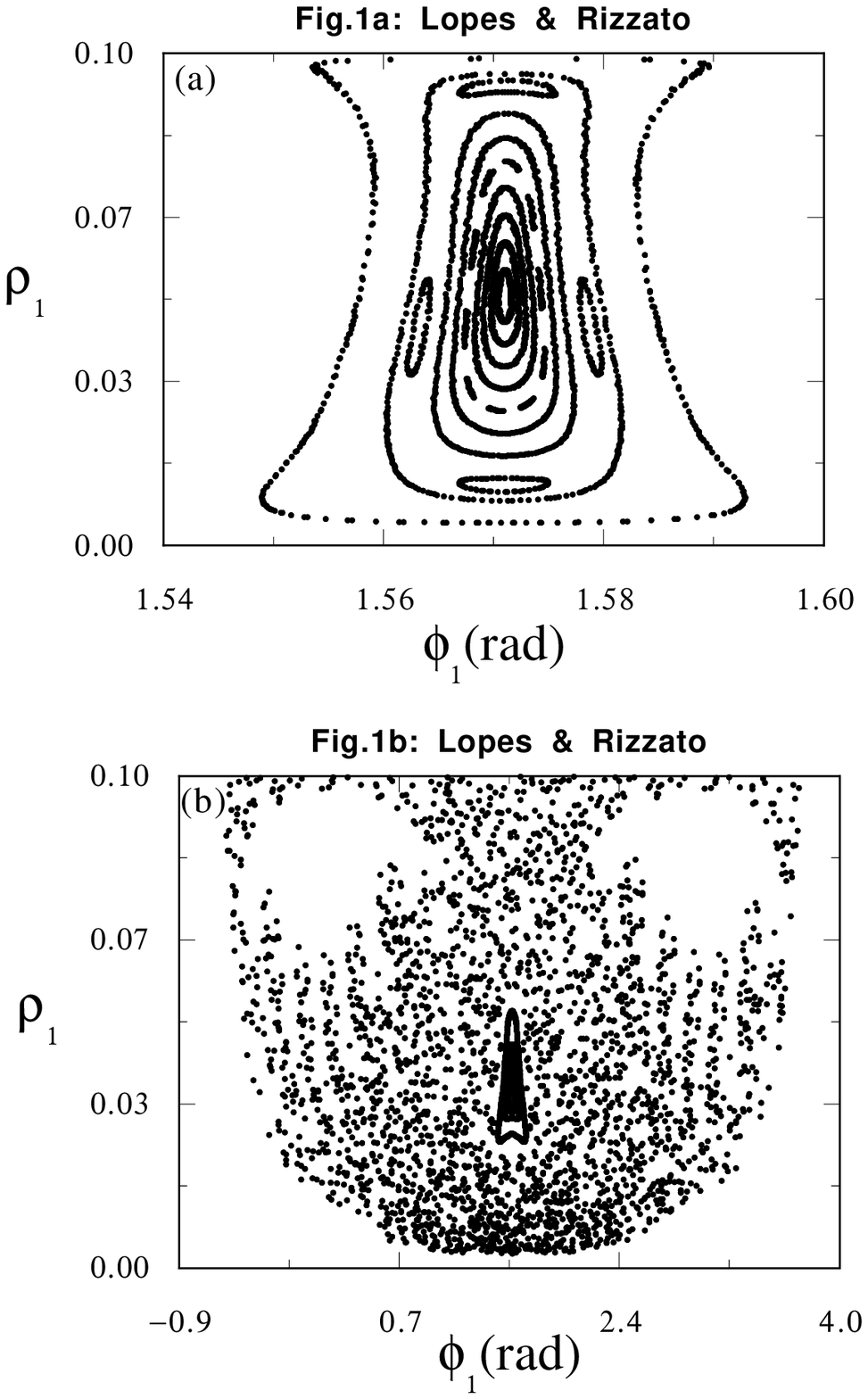
\label{fig1.ps}
\end{figure}

\begin{figure}
\caption{Time series of $\sqrt{<N^2>}$ for $r=1.0,\,v_{g_2}=1.0,\,N=128$, 
and various values of $\delta_3$. Initial conditions discussed in the text.}
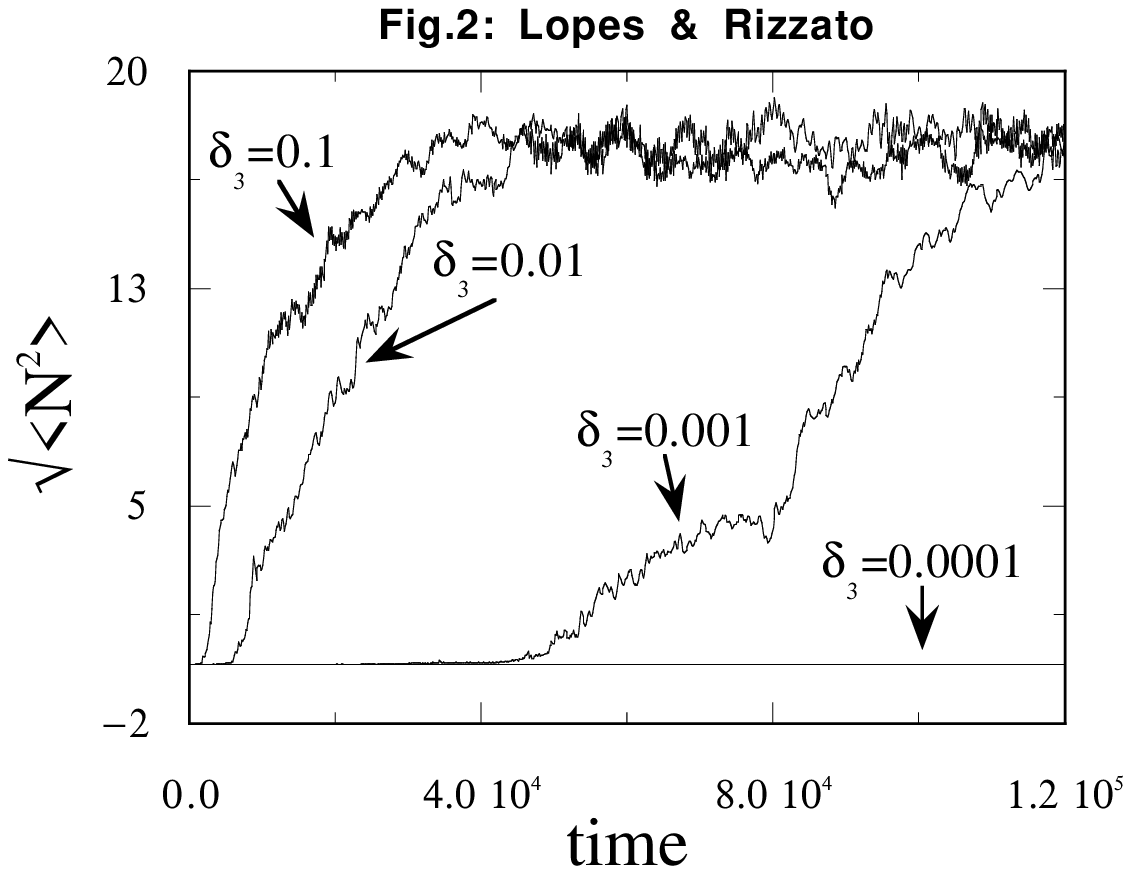
\label{fig2.ps}
\end{figure}

Let us consider as a first instance the case of a chaotic homogeneous 
manifold combined with linearly stable perturbations. To simulate this 
situation, we use the same parameters and techniques employed to 
construct Fig. (\ref{fig1.ps}b). In the homogeneous manifold the 
initial condition corresponds to the outermost orbit present in the figure. 
Since we had launched $10$ equally spaced orbits, a rapid calculation therefore 
shows that the value of $\rho_1 (t=0)$ to be used in the simulation reads 
$\rho_1 (t=0) = 0.09$. Fig. (\ref{fig1.ps}b) also shows that we take 
$\phi_1(t=0)=\pi/2$ for the initial phase. In addition we recall that $\phi_4=0$ 
and $h=0$. Note that the relations between $\rho$'s, $\phi$'s and $a_{j_o}$ are 
the same as previously defined in Eq. (\ref{introduzro}). For the 
inhomogeneous perturbation, besides $k=1.0$, we take $v_{g_2}=1.0$ as the 
relevant group velocity. The perturbing field itself is 
written in the form $Re[a_{1_1}]=|a_{1_0}| \times 10^{-3}$, 
$Im[a_{1_1}]=|a_{1_0}| \times 10^{-3}$ where $Re[\bullet]$ and 
$Im[\bullet]$ respectively denote real and imaginary 
parts of the various fields. For this particular choice of parameters and 
initial conditions, the inhomogeneous perturbations alone would be linearly 
stable, since here we have $k_{max} \approx 0.4 < k = 1.0$. On the other hand 
the low-dimensional phase-space is chaotic, and because of this fact, some kind 
of energy transfer should be expected after all. The initial results of our 
simulations can be found in Fig. (\ref{fig2.ps}). It is clearly seen that in 
spite of the linear stability of the system, transfer does indeed occur as 
$\sqrt{<N^2>}$ continuously grow until it reaches saturation due to the finite 
number of modes used in the numerical code. What 
happens here can be understood in terms of the stochastic pump model mentioned 
before, and discussed, for instance, in the book of Lichtenberg \& 
Lieberman \cite{lili83}. The strong chaotic dynamics developing at $k=0$ acts 
like a random force, diffusively driving all the remaining modes of the system. 
It does not matter if these remaining modes are linearly stable. What really 
matters is that as a result of the stochastic pump the inhomogeneous modes 
start to be driven similarly to what would happen if they were in contact with 
a thermal reservoir. Simulations with other values of $\delta_3$ are also shown 
in Fig. (\ref{fig2.ps}). It is seen that as one lowers $\delta_3$, 
redistribution weakens. In the limit of very small values of the mismatch, one 
recovers the situation depicted in Fig. (\ref{fig1.ps}a) where the homogeneous 
manifold is regular. Then diffusion apparently ceases or becomes very slow, 
what leads us to draw a most expected conclusion on the features of the system: 
when the homogeneous manifold is regular and no harmonic mode falls within 
the linear instability band, diffusion is negligible. 

\begin{figure}
\caption{(a) Comparison of time series of $\sqrt{<N^2>}$ for different 
number of modes used in the simulations; (b) Spectral-energy distribution 
of the case $N=256$ above, averaged over time and over the mode index $j$. 
In all cases $\delta_3=0.1,\,r=1.0,\,v_{g_2}=1.0$; initial conditions 
discussed in the text.}
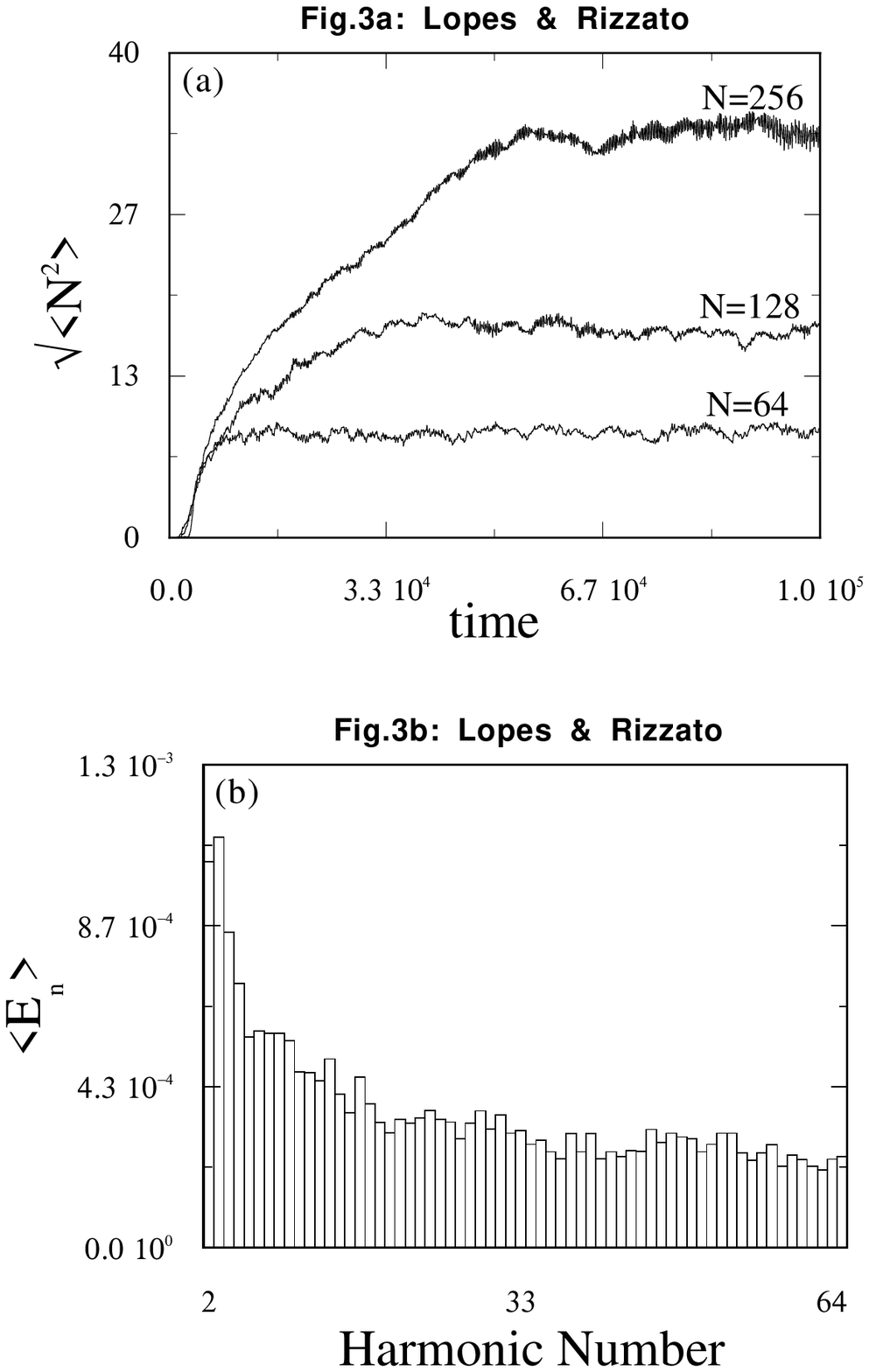
\label{fig3.ps}
\end{figure}

Figure (\ref{fig3.ps}a) shows that the larger the number of modes used in the 
simulations, the larger the saturated value of $\sqrt{<N^2>}$. In other 
words there is no nonlinear saturating effect that involves only a small and 
fixed number of modes. In addition to that, the various numbers of modes used 
in the figure, $N=64,\,128$, and $256$, indicate that diffusion is not an 
effect connected with the finite discretization of the numerical method. 
Diffusion is present no matter the value of $N$. One can see this because 
for earlier times of the nonlinear interaction where a few modes are 
actually excited, the behavior of all curves are virtually the same.   
A close inspection of the final spectral-energy distribution reveals that our 
asymptotic states are not quite fully equipartitioned in energy. Although the 
spectral distribution for large mode numbers is relatively flat, the 
energetic content in the spectral region of small mode numbers is slightly 
larger. This can be seen with help of Fig. (\ref{fig3.ps}b) where we plot 
the energy of the free-modes, $E_n = \ll(\delta_j+n k )\,|a_{j_n}|^2\gg$, 
doubly averaged over time and over the wave index $\{j=1,2,3,4\}$. 
This discrepancy between full equipartition and simulations could have two 
contrasting sources that should be investigated: either the wave amplitudes 
we work with are already so high that the interaction term is no longer a small 
perturbation, or for the parameters used here integrable features are still 
appreciable. It is perhaps convenient to mention that the convergence of the 
spectral distribution has been verified by extending and reducing the averaging 
time interval in various ways; for sufficiently long intervals the results 
are the same. 

\begin{figure}
\caption{Space-time history of $|A_1(x,t)|$ for the $\delta_3=0.1$ case 
of Fig. (2). By the end of the simulation inhomogeneities are already 
strongly excited.}
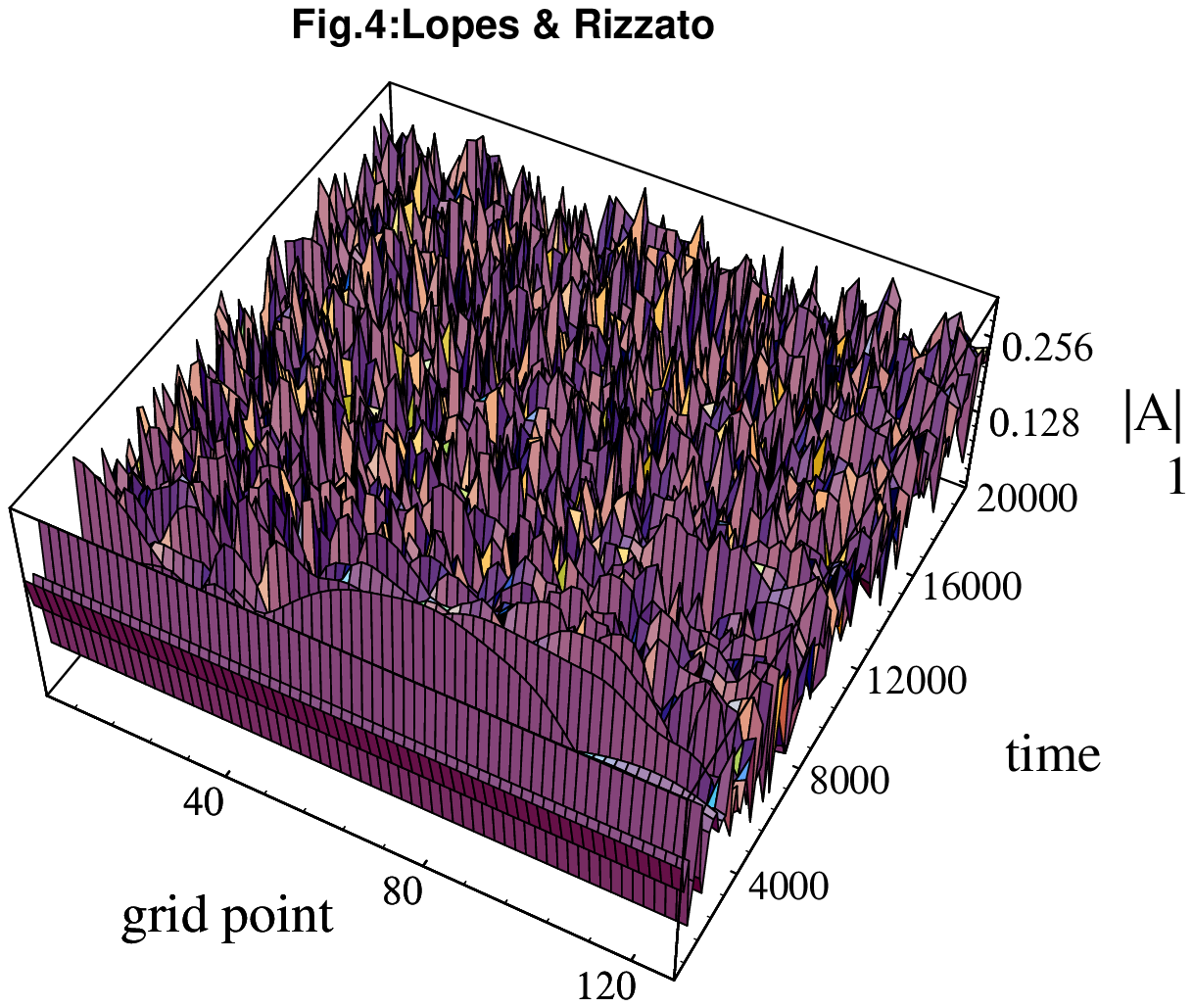
\label{fig4.ps}
\end{figure}

To complement the information obtained with help of Figs. (\ref{fig2.ps}) and 
(\ref{fig3.ps}), in Fig. (\ref{fig4.ps}) we display the space-time history of 
$|A_1(x,t)|$ corresponding to the $\delta_3=0.1$ curve of Fig. (\ref{fig2.ps}). 
It may be seen that for earlier times of the nonlinear interaction, one has a 
purely time dependent solution that is completely homogeneous in space. 
However, as time is let to evolve, strong spatial oscillations adds to the 
previous erratic but pure time dependence. The upper time limit of the 
simulation is $t=20000$ which is smaller than the asymptotic time where the 
system reaches saturation. Nevertheless, it is clear from the figure that 
inhomogeneities are already strongly excited by then. 

\subsection{The $\delta_3 = 0$ case.}

\begin{figure}
\caption{Time series of $\sqrt{<N^2>}$ when orbits of the regular type 
evolve on the homogeneous manifold. The number of linearly 
unstable modes in all cases is approximately $5$. In all cases 
$\delta_3=0$, $v_{g_2}=0.06$ and $N=128$; initial conditions discussed 
in the text.}
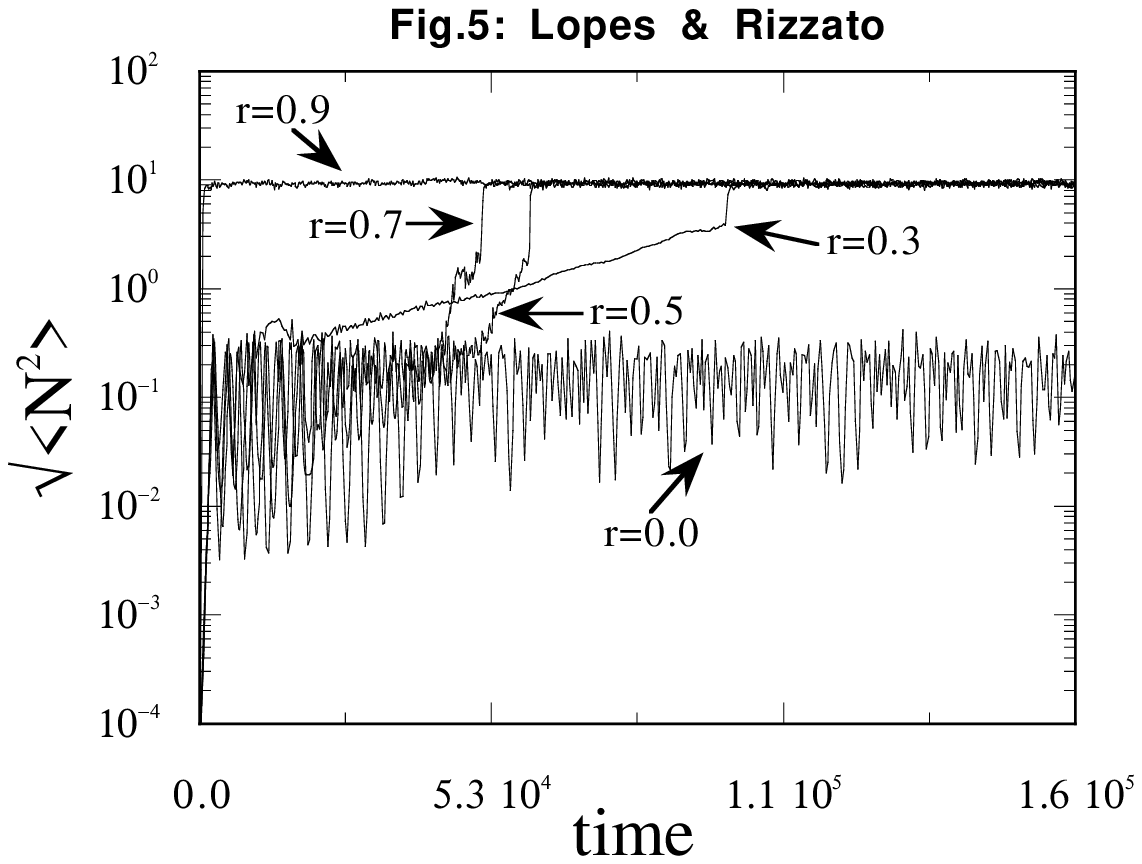
\label{fig5.ps}
\end{figure}

Now let us consider the other individually relevant situation in which we 
combine a regular manifold with a linearly unstable perturbation. The relevant 
graphical information can be found in Fig. (\ref{fig5.ps}). In this case we 
work with $\delta_3=0$ so as to guarantee that the homogeneous manifold is 
in fact regular. The choice of initial conditions follows the pattern used to 
construct the previous figures. We launch $10$ initial conditions on the 
homogeneous manifold equally spaced between $\rho_1=0$ and $\rho_{max}=0.1$, 
and select the initial condition for the full simulations as the one 
corresponding to the outermost orbit with $\phi_1=\pi/2$. We choose 
$v_{g_2}=0.06$ such as to have at least some linearly unstable modes. 
In fact, within the range of variation of $r$ in the figure, use of 
Eq. (\ref{reldis}) shows that the number of unstable modes is approximately 
$5$ for all cases. 

One could think that as chaotic activity is absent from the homogeneous 
manifold, no diffusion would be observed now. This proves to be 
untrue. Indeed, Fig. (\ref{fig5.ps}) shows that when $r \neq 0$, energy 
diffusion is {\it again} present, even though homogeneous chaos is 
not (remember that $\delta_3=0$). We had seen before that for coupled 
triplets - that is if $r \neq 0$ - purely temporal chaos alone is capable to 
drive energy redistribution, even though inhomogeneous perturbations are 
linearly stable. Now what we are observing is a complementary effect. Namely, 
unstable inhomogeneous modes alone are capable to drive redistribution if 
$r \neq 0$ even though the homogeneous manifold is integrable ($\delta_3=0$). 
In this case the active modes cannot be excited with the stochastic pump 
simply because there is no stochastic pump available. Instead they are excited 
by the linear instability of the system, and as soon as they are excited 
diffusion starts to operate again. This feature leads us to reconsider the 
first case analyzed here. In fact, in this first case where we have a chaotic 
homogeneous manifold combined with stable spatial modes diffusion can be 
understood in terms of the stochastic pump model only for earlier times of the 
nonlinear interaction, when the inhomogeneity is relatively small. Only then 
are the effects of the spatial inhomogeneities small, with diffusion totally 
dictated by the homogeneous chaos. As soon as the inhomogeneous fluctuations
grow larger, both nonintegrable effects become of the same order of magnitude.
At this stage the overall dynamics should become similar to that case where
one combines a chaotic manifold with unstable inhomogeneous modes, a case not
individually explored in this work.

\section{Final Remarks}

A general comment is that in the wide sense investigated here, the 
global stability of the homogeneous manifold cannot be 
simply understood in terms of a linear stability analysis. In fact, 
we have shown explicitly that the linear analysis fails when the 
homogeneous manifold is chaotic. To emphasize our point we have analyzed 
a particular setting where the inhomogeneous modes that we have launched 
are linearly stable. Even so, inhomogeneous perturbations tend 
to grow as time evolves. What happens in this case is that at least for 
earlier times of the nonlinear interaction, the chaotic 
homogeneous manifold acts like a stochastic pump delivering 
energy in a random fashion to these inhomogeneous modes. One calls 
that situation the stochastic pump because the role of the 
homogeneous manifold is comparable to the role of a thermal reservoir 
diffusively delivering energy to its vicinities \cite{lili83,goe92}.

When homogeneous chaos is absent, two types of behaviors could 
be identified for $r \neq 0$. 
If no inhomogeneous perturbation falls within the band of 
linear instability, then the system does not develop any sort of 
configuration with spatial dependence and simply keeps supporting 
a homogeneous periodic dynamics. On the other hand, when at least 
one of the inhomogeneous modes does fall within the instability 
band one has again diffusion towards small spatial scales. This 
shows that the nonintegrable features of our system are not 
exclusively related to the temporal dynamics developing on the homogeneous 
manifold. In the homogeneous manifold, chaos vanishes as 
$\delta_3 \rightarrow 0$, but for dynamical states involving unstable 
modes with $k \neq 0$, energy redistribution is present irrespectively of 
the value of $\delta_3$. In other words, even if $\delta_3 = 0$ in these 
situations, random activity is still present. 

The spatio-temporal dynamics of one isolated wave triplet has been shown to be 
integrable via inverse-scattering \cite{kau79,chow95}. Our results indicate that 
as one couples a fourth wave such as to form two wave triplets sharing two 
common modes, integrability breaks down. What seems to be happening here 
is that as one couples the fourth mode, the regular solutions of the isolated 
triplet undergo a transition to chaos. 
Then, the resulting chaotic dynamics enhances radiative effects and causes 
energy transfer to 
modes with larger and larger wavevectors; this and the following are issues under 
current investigation. The regular asymptotic dynamics of the pure triplet is 
somewhat related to the intermediary dynamical stages of the four-wave system. 
Consider, for instance, an ordering of group velocities for which solitons 
saturate the pure triplet interaction, as commented earlier. When 
the fourth-wave is coupled to the unperturbed triplet, integrability 
is destroyed and solitons survive only for finite amounts of time, ultimately 
releasing their energy into small scale fluctuations. In other words, regular 
structures that are the final asymptotic states of the triplet interaction, 
are present only during intermediary stages of the full four-wave system. 
In addition, although the ordering of the group velocity is relevant 
during intermediary stages of the full interaction, we see that it is not 
expected to drastically affect the asymptotic dynamics and the overall 
process of energy transfer: eventually modes with short spatial scales 
are excited and intermediary regular structures are no longer present in 
the system. 

Another important point to be considered is that the 
general technique of multiple time scales leading to generic amplitude equations 
breaks down when modes with large enough wavevectors become involved in the 
dynamics. Then, dispersive and dissipative effects should be included since 
these terms are always related to small wavelengths - dispersion because 
it appears in the form of derivatives terms like $\partial^2 A_j / \partial x^2$ 
such that the smaller the length scale, the larger the derivative, 
and dissipation because it occurs when the waves strongly interact with small 
scale structures of the global system, like for example high-velocity 
particles in plasmas. While we are currently considering these points, we emphasize 
that our present model works well for earlier times of energy transfer when modes 
with small wavelengths are still of very small amplitudes. In particular, all the 
conditions analyzed here that engage transfer, involve only large wavelengths and are 
therefore fairly accurate.

\centerline{\bf ACKNOWLEDGMENTS}

We would like to thank nice discussions with R. Pakter, A. Chian, 
and R. Viana. We also thank suggestions from two anonymous referees. This 
work was partially supported by Financiadora de Estudos e Projetos (FINEP), 
Conselho Nacional de Desenvolvimento Cient\'{\i}fico e Tecnol\'ogico (CNPq), 
and Funda\c{c}\~ao da Universidade Federal do Paran\'a (FUNPAR), Brazil. 
Numerical computing was performed on the CRAY Y-MP2E at the Universidade 
Federal do Rio Grande do Sul Supercomputing Center.  

\newpage

\newpage
%
%
%
%
%
%

\begin{thebibliography}{99}
%
\bibitem{kau79}D.J. Kaup, A.H. Reiman, and A. Bers, Rev. Mod. Phys. {\bf 51} 
(1979) 275.
%
\bibitem{wil77}J. Weiland and H. Wilhelmsson, {\it ``Coherent Non-Linear 
Interaction of Waves in Plasmas''}, Pergamon Press (1977).
%
\bibitem{chow95}C.C. Chow, Physica D {\bf 81} (1995) 237.
%
\bibitem{sug68}R. Sugihara, Phys. Fluids {\bf 11} (1968) 178.
%
\bibitem{kar73}K.S.Karplyuk, V.N. Oraevskii, and V.P. Pavlenko, Plasma 
Phys. {\bf 15} (1973) 113.
%
\bibitem{wal77}D. Walters and G.J. Lewak, J. Plasma Phys. {\bf 18} 
(1977) 525.
%
\bibitem{chi96}A.C.-L. Chian, S.R. Lopes, and J.R. Abalde, Physica D, 
{\bf 99} (1996) 269.
%
\bibitem{pak97}R. Pakter, S.R. Lopes, and R. Viana {\it ``Transition to 
Chaos in the Conservative Four-Wave Parametric Interactions''} Physica D 
{\it accepted} (1997).
%
\bibitem{lili83}A.J. Lichtenberg and M.A. Lieberman, {\it ``Regular and 
Stochastic Motion''}, Springer (1983).
%
\bibitem{pet90}M. Pettini and M. Landolfi, Phys. Rev. A {\bf 41} (1990) 
768.
%
\bibitem{goe92}C.G. Goedde, A.J. Lichtenberg, and M.A. Lieberman, 
Physica D {\bf 59} (1992) 200.
%
\bibitem{tsa96}G. Tsaur and J. Wang, Phys. Rev. E {\bf 54} (1996) 4657. 
%
\bibitem{riz92}F.B. Rizzato and A.C.-L. Chian, J. Plasma Phys. {\bf 48} (1992) 71; 
R. Pakter, R.S. Schneider, and F.B. Rizzato, Phys. Rev. E {\bf 49} (1994) 1594.
%
\bibitem{nos97}G.I. de Oliveira, L.P.L. de Oliveira, and F.B. Rizzato, 
Physica D {\bf 104} (1997) 119.
%
\bibitem{rom83}F.J. Romeiras, Phys. Lett. {\bf 93A} (1983) 227.
%
\bibitem{yue80}D.U. Martin and H.C. Yuen, Phys. Fluids {\bf 23} (1980) 1269.
%
\bibitem{thy81}A. Thyagaraja, Phys. Fluids {\bf 24} (1981) 1973.
%
%
\end{thebibliography}
\end{document}